# Justice in Healthcare Artificial Intelligence in Africa

Aloysius Ochasi[a], Abdoul Jalil Djiberou Mahamadou[b], Russ B. Altman[c, d, e, f]

[a]Department of Bioethics and Interdisciplinary Studies, Brody School of Medicine, East Carolina University, Greenville, NC 27858, USA
[b]Stanford Center for Biomedical Ethics, Stanford University, Stanford, CA 94305, USA,
[c]Department of Biomedical Data Science, Stanford University, Stanford, CA 94305, USA
[d]Department of Bioengineering, Stanford University, Stanford, CA 94305, USA
[e]Department of Genetics, Stanford University, Stanford, CA 94305, USA
[f]Department of Medicine, Stanford University, Stanford, CA 94305, USA

**Abstract**
There is an ongoing debate on balancing the benefits and risks of artificial intelligence (AI) as AI is becoming critical to improving healthcare delivery and patient outcomes. Such improvements are essential in resource-constrained settings where millions lack access to adequate healthcare services, such as in Africa. AI in such a context can potentially improve the effectiveness, efficiency, and accessibility of healthcare services. Nevertheless, the development and use of AI-driven healthcare systems raise numerous ethical, legal, and socio-economic issues. Justice is a major concern in AI that has implications for amplifying social inequities. This paper discusses these implications and related justice concepts such as solidarity, Common Good, sustainability, AI bias, and fairness. For Africa to effectively benefit from AI, these principles should align with the local context while balancing the risks. Compared to mainstream ethical debates on justice, this perspective offers context-specific considerations for equitable healthcare AI development in Africa.

**Introduction**
Artificial intelligence (AI) driven by big data is fueling the fourth industrial revolution. It can potentially revolutionize countless aspects of our lives and transform the healthcare industry in unprecedented ways. The concept of AI, which can be traced back to the 1950s, is "an umbrella term for a range of techniques that can be used to make machines complete tasks in a way that would be considered intelligent were they to be completed by a human (Morley et al. 2020). More specifically, AI in the healthcare milieu refers to "the use of intelligent data-driven technologies that leverage healthcare resources and data more effectively to support and streamline decision-making in healthcare and to consequently provide better healthcare services that are tailored to individual needs"(Siala and Wang 2022). In the Global South, particularly in Africa, where millions have inadequate access to essential health services (WHO 2020) due to the underdevelopment of healthcare infrastructures and limited resources, AI has the potential to significantly improve the efficiency, effectiveness, and accessibility of healthcare services. This includes the delivery of healthcare services in rural areas (Guo and Li 2018), healthcare management (Schwalbe and Wahl 2020), disease outbreak prediction and forecasting (Mbunge and Batani 2023), and medical diagnosis (Owoyemi et al. 2020). However, the successful integration of healthcare AI in Africa requires addressing several ethical, legal, and socio-economic challenges, including implications arising from data collection, use, and sharing, transparency, accountability and responsibility, and social justice.

The principle of justice entails promoting prosperity, preserving solidarity, and avoiding unfairness (Floridi and Cowls 2019) to ensure that "no person or group is subject to discrimination, neglect, manipulation, domination, or abuse" (WHO 2021). Solidarity, Common Good, and sustainability are closely allied with justice, which must be addressed to harness the benefits of AI in Africa and minimize the risks. Solidarity in the context of AI-driven healthcare involves a collective commitment to ensuring that the benefits of technological advancements are equitably distributed and accessible to all individuals regardless of their geographic location or socioeconomic status. Such commitment also calls for sustainable use of resources such as energy and water, which are already scarce in Africa and needed for AI development. A failure in the sustainable development of AI could deepen existing social inequalities across the continent.

In this paper, we argue that justice and these related concepts are integral to the successful integration of AI in Africa and essential for the development and use of AI in healthcare that aligns with the local context. We propose a moral framework as an alternative to mainstream debates on AI justice dominated by Western countries (Jobin, Ienca, and Vayena 2019). In the remaining paper, we first review existing ethical AI frameworks for healthcare in low-resource countries followed by discussions on (distributive) justice, the need for equitable AI access, the principles of the Common Good, solidarity, and sustainability. Finally, we discuss AI bias and fairness focusing on sources of bias in an African context.

1. **Beyond Bias and Fairness**

There is a growing effort in the scholarly and grey literature to contextualize mainstream healthcare ethical AI frameworks criticized for being Western-centric (Eke, Wakunuma, and Akintoye 2023; Jobin, Ienca, and Vayena 2019). Fletcher and colleagues (Fletcher, Nakeshimana, and Olubeko 2021) discussed bias, fairness, and the appropriate use of AI in global health and provided guidelines and recommendations for AI appropriateness, bias identification, and fairness enforcement. In a similar context, (Kong et al. 2023) presented the Responsible, Explainable, and Local Artificial Intelligence for Clinical Public and Global Health in the Global South (REL-AI4GS), an ethical AI framework for clinal public and global health with key components on responsibility, explainability, and locality. Furthermore, (Wahl et al. 2018) explored the issues that could undermine patients' and clinicians' trust in the accuracy of AI systems, data access, data ownership, AI bias and transparency, and the "Do no harm" principle implications in the context of AI and global health. Alami and colleagues (Alami et al. 2020) proposed five building blocks for responsible, sustainable, and inclusive healthcare AI in low-resource countries: the training and retention of local expertise, robust system monitoring, system-based approaches for the implementation of effective and reliable healthcare AI systems, and inclusive local actors and stakeholders including women and minority, an poor communities in the development of AI. Other scholarly works have focused on the contextualization of data access, privacy, and protection (Sallstrom, Morris, and Mehta 2019), data availability and quality (Owoyemi et al. 2020), accountability (Sallstrom, Morris, and Mehta 2019), infrastructure inadequacy (Mbunge and Batani 2023; Owoyemi et al. 2020), and cost of access to AI (Owoyemi et al. 2020). From a technical lens, in a series of qualitative and quantitative studies, Asiedu and colleagues discussed African perceptions of AI bias and fairness and guidelines for algorithmic fairness attribute selection (Asiedu et al. 2024; 2023).

Justice in these frameworks is often discussed from AI bias and fairness perspective (Asiedu et al. 2024; 2023; Fletcher, Nakeshimana, and Olubeko 2021; Wahl et al. 2018) and technological access (Owoyemi et al. 2020). Alami and colleagues discussed sustainability in the context of multi-stakeholder partnerships and technology as a means to achieve sustainable communities (Alami et al. 2020). Here, we argue for sustainable AI-driven healthcare systems to prevent the amplification of social inequities by empowering equitable access to energy and water and reducing the demands of these resources in AI development. Moreover, we argue for technology solidarity and technology for the common good.

## 2. Bridging the Justice Gap

Justice is a complex concept that has been analyzed and interpreted through various philosophical lenses in human civilization. It has been referred to as the "proverbial elephant" examined by six blind individuals where each feels and describes a different part of the elephant as the real elephant. In his Nicomachean Ethics, Aristotle views justice as a virtue essential for human flourishing and societal well-being, which consists of what is lawful and fair. Fairness, for him, essentially requires equitable distribution and the correction of what is inequitable. Consequently, distributive justice involves dividing benefits and burdens fairly among members of a community. On the other hand, corrective justice requires, in some circumstances, the restoration of a fair balance in interpersonal relations where it has been lost (Irwin 2019). In contrast, John Rawls introduced a contemporary approach to distributive justice, termed "justice as fairness" and proposes a systematic approach to distributive justice designed to ensure that social and economic inequalities are arranged to the maximum benefit of the least advantaged members of society (Rawls 1971). The limited scope of this paper will not permit an in-depth exploration of the concept of justice, but suffice it to say that justice will be understood as a principle that dictates the equitable and fair distribution of benefits and burdens among members of society throughout this paper.

The deployment of healthcare AI in Africa raises issues of justice. Justice recognizes that each person should be treated fairly and equitably and be given his or her due. The issue of medical disparities among the wealthy and poor nations focuses on distributive justice: the fair, equitable, and appropriate distribution of medical resources in society. Distributive justice requires that everyone receive equitable access to the primary health care necessary for living a fully human life insofar as there is a fundamental human right to health care (Ochasi and Clark 2015). Africa is a continent of 1.4 billion people (about 18% of the global population) that contributes less than three percent of the global GDP but carries over 20% of the global burden of disease (Niohuru 2023). The disparity in healthcare between the developed nations and lower to middle-income countries in the Global South is glaring, especially the digital divide in healthcare infrastructure essential for AI integration in Africa (Kondo et al. 2023). In the Global North, especially the United States (US) and Europe, healthcare AI is already revolutionizing the delivery of care in various spheres of medicine, such as diagnostics, personalized medicine, and the pharmaceutical industry (Sharma et al. 2018). The integration of AI in healthcare in Africa faces a significant challenge regarding the substantial financial investment required to develop, implement, and sustain complex and costly technologies for delivering digital healthcare services effectively (Chengoden et al. 2023). The initial high costs of implementing and maintaining these technologies may strain and collapse Africa's meager financial resources for healthcare practices and systems, potentially hindering their widespread adoption.

Justice demands that we bridge the divide in the lopsided deployment of healthcare AI through capacity-building in Africa. Most of the advances in AI and data generated in Africa are owned by private equity firms, corporations, multinationals, and international organizations whose infrastructure is developed outside the continent aided by African AI experts who work for them because of the lack of substantive and significant AI R&D in Africa (Gwagwa et al. 2020; Ndiaye 2024). The Montreal Declaration strongly argues that "the development of AI should promote justice and seek to eliminate all types of discrimination." Similarly, the European Commission's Group on Ethics in Science and Technology asserts that AI should "contribute to global justice and equal access to the benefits" of AI technologies (Floridi and Cowls 2019). Therefore, it is a moral imperative that all AI stakeholders invest in building the capacity for healthcare AI development and implementation in the continent. They can start by investing in young African entrepreneurs, local developers, policymakers, and healthcare workers interested in AI development and application in healthcare. Such an initiative would create room for local solutions and change the ugly narrative that foreign AI companies use false African identities (and experts) as marketing tools to raise capital and eventually cash out (Pilling 2019). It would also assuage the fears of data colonialism and exploitation in an era when the data revolution in Africa is described as the 'new gold' or the 'new oil' (Keymanthri Moodley and Rennie 2023). Building local capacity for healthcare AI development would expand access to healthcare so that those in most need of care would have access to it. Ultimately, the unequal distribution of healthcare AI has the propensity to exacerbate health inequities in Africa, violating distributive justice (Corbett-Davies et al. 2017). Similarly, inequitable access to healthcare AI can deepen existing health and socio-economic divides.

   3. **Ensuring Equitable Access to AI Technologies**

The principle of justice emphasizes fairness in the distribution of social goods. However, when considering AI as a social good, there are wide geographical disparities within and between countries, notably in access to technology. A significant contributing factor is the high cost of AI development and implementation, varying from a few hundred to several millions of dollars (Schwartz et al. 2020). Consequently, high-resourced countries with ample financial resources can more easily invest in the technology. In contrast, low-resourced countries with scarce funding could have limited access to the technology and benefits. Besides the high-resourced and low-resourced divide, a direct consequence of inequitable access to technologies is the urban and rural divide in Africa. In addition to the technological divide, limited technological access can amplify socioeconomic inequalities and have substantive cultural consequences by reinforcing stereotypes, cultural representation, and visibility (Cacal 2024). To address these issues, WHO calls for "industry and governments should strive to ensure that the 'digital divide' within and between countries is not widened and ensure equitable access to novel AI technologies" (WHO 2021), a call endorsed by UNESCO (UNESCO 2021).

Under limited funding opportunities, African countries can benefit from state-of-the-art open-sourced AI technologies. Open-source AI has been advocated to enhance transparency, foster collaboration, stimulate innovation, and increase the accessibility of AI technologies. Open-source platforms, such as [Hugging Face](), provide diverse pre-trained AI models and resources that can be adapted and finetuned to suit specific needs and contexts, particularly in Africa. Countries with limited expertise and resources can access cutting-edge AI tools without the burden of excessive costs typically associated with proprietary models. However, these opportunities could come with additional financial burdens and ethical implications.

Deploying open-sourced AI models may require adapted infrastructures (e.g., computing resources) to effectively work, these resources are often scarce in African countries and pose a significant challenge to unlocking AI's full potential. Furthermore, open-sourced AI models can inherit biases present in the data used to train the models, such as the overrepresentation of individuals of European descent (Chen et al. 2023). This can lead to bias and unfairness when unaddressed and deployed in Africa. Similarly, these models raise concerns about privacy if sensitive data were used to train the models along with vulnerability and security threats.

Disruptive technology development, such as AI, should be driven by the Common Good to benefit society.

### 4. Prioritizing the Common Good Over Corporate Greed

Social justice is connected with the Common Good. The adoption of healthcare AI in Africa must be driven by concern for the Common Good, not corporate greed. The Common Good is defined as "certain general conditions that are … equally to everyone's advantage." (Rawls 1971). It also consists of "our shared values about what we owe one another as citizens who are bound together in the same society" (Reich 2019). Though Reich narrowly used the term 'society' to refer to American society; we use it broadly to refer to human society and our moral obligations to one another as citizens of the human society (Reich 2019). The Common Good of the people in Africa could be described as the totality of socio-political, economic, religious, and cultural factors that help the individual flourish and realize her societal potential (Ochasi 2017). The disease burden in Africa endangers the Common Good, which is interwoven with the good life of the citizens. As Aristotle noted in his *Nicomachean Ethics*, "A good life is oriented to goods shared with others-- the common good of the larger society of which one is part. The good life of a single person and the quality of the common life persons share in society are linked. Thus, the good of the individual and the common good are inseparable" (Hollenbach 2002).

The deployment of healthcare AI on the African continent offers excellent opportunities to enhance the Common Good by expanding and improving access to healthcare through remote consultations and diagnoses, thereby helping healthcare practitioners do more with limited resources (Wahl et al. 2018; Lannquist 2021). The WHO estimates that by 2030, there will be a shortage of 18 million healthcare workers, predominantly in lower to middle-income countries (WHO, n.d.)AI has the potential to bridge the gap and bring relief to millions of Africans who have difficulty accessing adequate healthcare due to costs, dilapidated health infrastructure, and overburdened healthcare systems (Lannquist 2021). AIUK and Asilomar AI Principles assert that AI should "be developed for the common good and the benefit of humanity" (Floridi and Cowls 2019). However, the proprietary nature of healthcare AI technology, which falls under the ownership of private companies, could lead private equity firms, corporations, and AI developers to prioritize profits over access, which not only imperils the Common Good but negates the benefits of AI in resource-poor settings (Crawford et al. 2016). Resource-rich countries and corporations with better access to extract more data from resource-poor countries in Africa at higher speeds should avoid undue monetization and commercialization of AI services created from this data; concern for the Common Good should elevate fairness over profit generation (K. Moodley 2023). Choosing corporate greed over corporate social responsibility in a continent grappling with poverty and overburdened by diseases from malaria, TB, and HIV/AIDS is a violation of the principle of the Common Good.

When the Common Good principle drives AI development, solidarity becomes a cornerstone of technological advancement.

### 5. Promoting Global Solidarity

AI is poised to usher in the fourth Industrial Revolution and is already improving healthcare delivery in many high-income countries. We now live in a 'global village' and are more interconnected and interdependent on one another than ever in human civilization. The concept of human solidarity refers to "a disposition that each can have to act in solidarity with some others…a willingness to acknowledge need in everyone else and to act in general ways to support their human rights, especially by working toward the construction of transnational institutions that can allow for their fulfillment worldwide…" (Gould 2007). Solidarity aligns with Ubuntu, a South African philosophical concept representing universal interdependence and communalism. It advocates for interactions that nurture sharing and building trusting relationships filled with mutual compassion and respect. It also requires listening and affirming others, sharing resources, and making essential services available (Nussbaum 2003; Gilliam 2021). The World Health Organization affirmed the right to health as a human right, which includes four essential elements: availability, accessibility, acceptability, and quality (WHO 2023). Human solidarity demands that we care about what is happening worldwide, especially in low- to middle-income countries. We argue that since individuals have a right to health care, solidarity calls for sustained collaboration and partnerships among AI stakeholders and local governments, NGOs, healthcare providers, academic research institutions, and civil society organizations to ensure that through sustained "localized and innovative interdisciplinary research," the benefits of AI are aligned with the healthcare needs of the continent (Gwagwa et al. 2020). One of the significant challenges in deploying healthcare AI in Africa is the lack of significant clinical and high-quality data sets for training AI models, which is directly related to high preparation costs and time associated with data acquisition (Owoyemi et al. 2020). Such impediments call for concerted efforts and resources to scale up the necessary infrastructure to build an ethical and responsible AI on the continent. The principle of solidarity demands that societal benefits that accrue from the deployment of healthcare AI should not be limited to high-income countries alone but should be available to those in most need of care by "making AI tools open-source and user-friendly" (Hamet and Tremblay 2017).

The development of healthcare AI in resource-constrained countries requires a sustainable approach to avoid exacerbating existing social inequities.

## 6. Ensuring Sustainable AI Development

There is growing interest in developing "local" AI technologies in Africa. However, the development of such technology requires large computing and infrastructure capacities, which can lead to substantial environmental impacts (Wu et al. 2022). For instance, modern generative AI models such as ChatGPT consume the energy of about 33,000 households (Crawford 2024). Besides energy consumption, AI impacts water consumption, with a global water demand estimated to be half of the United Kingdom to maintain optimal temperatures of servers and computing equipment (Crawford, 2024). These resource demands, even when relatively low compared to Western settings, can amplify existing energy and water resource scarcity in Africa. In fact, about two-thirds of the African population does not have access to electricity (Mukhtar et al. 2023), especially in Sub-Saharan Africa, where an estimated 40% of people lacked access to safe and affordable water for domestic use in 2017 (Leal Filho et al. 2022). In addition to the ecological impact, AI resource demands can increase social inequalities in Africa as access to energy is closely connected to safe water access, poverty, and income distribution (Acheampong, Dzator, and Shahbaz 2021; Sarkodie and Adams 2020).

Beyond the development of AI, sustainability should be addressed during the entire life cycle of AI (van Wynsberghe 2021). For instance, in low-resource settings, the lack of maintenance of new health systems, including AI-driven can lead to the waste of resources (WHO 2021). The prospects and potential benefits of developing healthcare AI in Africa should be balanced against the burdens it brings upon a continent already grappling with famine, water scarcity, deforestation, and a host of other environmental issues. According to the Precautionary Principle, the burden of proof falls upon AI stakeholders ranging from developers, researchers, and organizations to governments that AI is designed and deployed in ways that do not harm people and the environment. The Precautionary Principle (aka, the "Precautionary Approach") originated from the German principle of foresight or Vorsorge. It was developed in the 1970s in the context of environmental law and policy and grew in popularity in European policy circles during the 1980s (Gilbert, Van Leeuwen, and Hakkinen 2009; Golding 2001). There are varied definitions of the Precautionary Principle; however, the description of the principle given at the Wingspread Conference in 1998 is germane to the need for sustainability in Healthcare AI in Africa: "When activity raises threats of harm to human health or the environment, precautionary measures should be taken even if some cause and effect relationships are not fully established scientifically. In this context, the proponent of an activity, rather than the public, should bear the burden of proof" (Salter 1988). Granted that the principle originated in the environmental field, its application has spread to other areas such as health protection, regulation of new biotechnologies, and applied ethics (Holm and Stokes 2011). We argue that this principle applies to the case for sustainability in the development and deployment of healthcare AI in Africa to minimize AI-induced harm, especially given the potential energy and water impact on the continent.

Sustainable AI also demands an inclusive approach (Alami et al. 2020) that addresses bias and unfairness to benefit every individual, group, community, and society.

## 7. Addressing Bias and Enforcing Fairness

According to Friedman and Nissenbaum (Friedman and Nissenbaum 1996), biased computer systems are those that "systematically and unfairly discriminate against certain individuals or groups of individuals in favor of others." With the integration of AI algorithms into various aspects of computer systems, the potential for bias amplification has become a concern. In healthcare, bias can result from different sources and model development stages including data biases, algorithmic biases, clinician interaction-related biases, and patient interaction-related biases (Ueda et al. 2024). However, the primary source of bias often traces back to the data used to train the algorithms, which may inherently reflect societal biases, historical disparities, or systemic inequalities present within healthcare systems (Mehrabi et al. 2021; Drukker et al. 2023). When AI systems are trained on biased data or the algorithmic design choices are biased this results in unfair outcomes. Despite the wide range of technical and practical solutions proposed to mitigate bias and enforce fairness, there is no consensus in the literature on effective approaches (Cary et al. 2023).

We present non-exhaustive sources of bias from an African perspective in the following. Specifically, we discuss gender bias, representation bias, aggregation bias, and synthetic data bias.

**Gender Bias:** Gender biases in AI algorithms reinforce gender stereotypes, and perpetuate gender inequities and discrimination against women (Manasi et al. 2022). Across Africa, women and girls face persistent disparities in almost every aspect of life (Ahmed and Sey 2020). In healthcare, for instance, access to essential services in reproductive and maternal health remains scarce to half of women and girls (Pons-Duran et al. 2019), and the likelihood of HIV among adolescent girls is three times higher than among boys of similar age (Oyebanji and Okereke 2023). While AI can help bridge the gap in access to healthcare, in particular for women and girls, it can also perpetuate such biases and exacerbate existing inequalities from the underrepresentation of these minority populations in the training data of AI systems. As a response, UNESCO calls for the insurance that "gender stereotyping and discriminatory biases are not translated into AI systems, and instead identify and proactively redress these" (UNESCO 2021).

**Representation Bias**: Representation bias occurs when the data reflect a disproportionate representation of population subgroups. Such bias may result from historical bias, population distribution skewness, sampling bias, selection bias, and self-selection bias (Shahbazi et al. 2023). In Africa, due to historically unequal access to healthcare, women and girls are likely to be underrepresented in biomedical databases, thus leading to representation bias, as discussed in the preceding.

**Aggregation Bias:** Aggregation bias occurs when population data are inappropriately combined. Such aggregation "masks critical within-group differences and disparities, limiting the health and social services fields' abilities to target their resources where most needed" (Kauh, Read, and Scheitler 2021). For instance, a study combining Asian and Pacific Islanders (API) data as one race category found that the life expectancy of API was superior to that of the White population (Taparra and Pellegrin 2022). However, a disaggregation of this category unveils poor life expectancy among Pacific Islanders. In the African context, data aggregation has a dual implication. On the one hand, aggregated data could help address data scarcity, and improve the statistical power of AI models. For instance, aggregating rich data from urban areas with scarce rural data could help deliver AI-driven health services in rural areas. On the other hand, as in the API example, such an aggregation could hide health disparities. Moreover, this aggregation could be a source of representation bias by overrepresenting men and underrepresenting women.

**Synthetic Data Bias**: Collecting new biomedical data, for instance in Africa, can be expensive and time-consuming. To address such an issue, technical solutions have been proposed to improve the performance and generalizability of AI models under low-data settings (Bansal, Sharma, and Kathuria 2022). These techniques include data augmentation, which refers to artificially augmenting AI training data. Typically, an AI model is trained on a large database to learn patterns in the data, afterwards, this model can be used to generate artificial data with similar patterns. In Africa, this process could help to balance women' and girls' representation in biomedical data. However, this process has several ethical implications. First, artificial data can make actual data more noisy which can affect the models' performance. Second, the synthetic data generation process can be biased when the model or technique used is biased (Drukker et al. 2023). In the preceding example, if the AI model used to generate synthetic data is biased (e.g., when an AI model trained on large biobank databases that are reflective of European descents is used to augment underrepresented groups data), it could result in biased synthetic data. Third, it remains an open question on how real and synthetic data should be weighed when training AI systems. Should the real data be outweighed, and why?

**Dataset Shift Bias**: Dataset shift occurs when AI models' training and testing/deployment data characteristics are different. When not carefully monitored, dataset shift can affect the models' performance and generalizability, which can have substantial consequences (Subbaswamy, Adams, and Saria 2021). In healthcare, a shift can occur under changes in technology (e.g., new types of data-acquisition devices), population and setting (e.g., a model deployed in a new clinical practice or on new clinical demographics), and behavior (e.g., changes in patient behavior) (Finlayson Samuel G. et al. 2021). In Africa, dataset shifts may occur when AI models are trained on non-local data, for instance when the models are trained on data from the West and deployed locally. Besides the impact on the performance and generalizability of AI models, dataset shifts can lead to unfair outcomes. A fair model in one setting (e.g., the West) can be unfair in another setting (e.g., in Africa) (Barrainkua et al. 2023). Despite the technical solutions that have been proposed to mitigate AI unfairness under changing environments, there is an ongoing debate on effective strategies (Barrainkua et al. 2023).

**Conclusion**
Navigating the ethical principle of justice is essential for equitable AI-driven healthcare systems development in Africa. Under such a resource-constrained setting, access to technology for everyone regardless of their geographical, socioeconomic, and demographic status is imperative. As a common good, the development of AI systems should balance the financial interests of private organizations and public interests. This requires technology solidarity and sustainability to limit existing technology and socioeconomic divides and address bias and unfairness. Most importantly it is essential that AI development and discussions around its implications be culturally sensitive to respect the needs, values, and norms of local communities and bring their voices to mainstream debates. This approach would ensure that Africa truly benefits from AI and delivers digital-enhanced healthcare services to improve patient outcomes.